\documentclass[12pt]{article}
\usepackage{amsmath}
\usepackage{amssymb}
\usepackage{slashed}
\usepackage[square,numbers,sectionbib,compress]{natbib}
\usepackage{graphicx}
\usepackage{breqn}
\usepackage{bbold}
\usepackage{xcolor}
\usepackage{float}
\usepackage{graphics, setspace}
\usepackage{framed}
\usepackage[export]{adjustbox}
\usepackage[colorlinks=true,urlcolor=black,anchorcolor=black,citecolor=black,filecolor=black,linkcolor=black,menucolor=black,linktocpage=true,pdfproducer=medialab,pdfa=true]{hyperref}
\usepackage[symbol]{footmisc}
\allowdisplaybreaks

\newcommand{\eps}{\varepsilon}

\newcommand{\D}{\mathcal{D}}

\usepackage{pslatex}
\usepackage[latin1]{inputenc}
\usepackage[T1]{fontenc}

\makeatletter
\DeclareRobustCommand{\cev}[1]{%
  {\mathpalette\do@cev{#1}}%
}
\newcommand{\do@cev}[2]{%
  \vbox{\offinterlineskip
    \sbox\z@{$\m@th#1 x$}%
    \ialign{##\cr
      \hidewidth\reflectbox{$\m@th#1\vec{}\mkern4mu$}\hidewidth\cr
      \noalign{\kern-\ht\z@}
      $\m@th#1#2$\cr
    }%
  }%
}
\makeatother

\begin{document}

\numberwithin{equation}{section}

\begin{titlepage}
\noindent
\hfill May 2024\\
\vspace{0.6cm}
\begin{center}
{\LARGE \bf 
    Anomalous dimensions for hard exclusive processes\footnote[1]{Presented at the XXXI International Workshop on Deep Inelastic Scattering and Related Subjects, Grenoble, France, 8-12 April 2024.}}\\ 
\vspace{1.4cm}

\large
S.~Van Thurenhout$^{\, a}$\\
\vspace{1.4cm}
\normalsize
{\it $^{\, a}$HUN-REN Wigner Research Centre for Physics, Konkoly-Thege Mikl\'os u. 29-33, 1121 Budapest, Hungary}\\
\vspace{1.4cm}

{\large \bf Abstract}
\vspace{-0.2cm}
\end{center}
We give an overview of recent developments in the computation of the anomalous dimension matrix of composite operators in non-forward kinematics. The elements of this matrix determine the scale dependence of non-perturbative parton distributions, such as GPDs, and hence constitute important input for phenomenological studies of exclusive processes like deeply-virtual Compton scattering. Particular emphasis will be put on a recently developed method that exploits consistency relations for the anomalous dimension matrix which follow from the renormalization structure of the operators.
\vspace*{0.3cm}
\end{titlepage}

\section{Introduction}
\label{sec:intro}
The description of hadronic structure is an important topic in contemporary particle physics. In particular, the full connection between hadronic properties and properties of the constituent partons is still not completely clear. For example, there is the long standing proton-spin puzzle \cite{Aidala:2012mv,Leader:2013jra,Deur:2018roz,Ji:2020ena,osti_1986091}, which poses the question of how the spin-1/2 of the proton can be explained in terms of the spins and orbital angular momenta of the quarks and gluons inside the proton. Valuable insight into this type of question can be gained by performing high-energy scattering experiments that can resolve the inner structure of the proton, or the hadron more generally. The cleanest channels, both from an experimental and a theoretical point of view, typically correspond to the scattering of elementary particles like electrons off the hadron. As the main interaction of interest corresponds to the strong one, which is very successfully described by quantum chromodynamics (QCD), we can exploit the rather nice property of factorization. The latter tells us that, when studying a scattering process that involves some hard scale, physical quantities such as the cross-section factorize into two main ingredients. The first ingredient is a function describing short-distance interactions. Typically this corresponds to some partonic amplitude, which can be calculated using perturbative methods. The second ingredient describes long-range effects of the interaction, and contains valuable information about partonic properties inside the hadron. Important examples include the standard parton distribution functions (PDFs) and the generalized parton distributions (GPDs). Unfortunately, these objects are non-perturbative and hence standard perturbative techniques are not applicable. First-principle information can nevertheless be obtained from lattice QCD analyses, see e.g.~\cite{Ji:2020ena,Alexandrou:2020sml,Wang:2021vqy,Alexandrou:2021bbo,Scapellato:2022mai,Alexandrou:2022dtc,Alexandrou:2020zbe,Alexandrou:2021lyf}.\newline

On a theoretical level, the non-perturbative parton distributions correspond to hadronic matrix elements of composite QCD operators. In particular, this implies that the scale dependence of the distributions is set by that of the operators. The latter is characterized by their anomalous dimensions, which can be computed perturbatively through the renormalization of the partonic operator matrix elements. While this is straightforward in forward kinematics\footnote[1]{At least for the flavor-non-singlet operators. For the flavor-singlet ones complications arise because of the appearance of non-gauge-invariant operators, see e.g.~\cite{Hamberg:1991qt,Matiounine:1998ky,Blumlein:2022ndg,Falcioni:2022fdm,Gehrmann:2023ksf,Falcioni:2024xyt}.}, relevant for inclusive scattering processes and PDFs, the extraction of the anomalous dimensions is somewhat obscured in non-forward kinematics, relevant for exclusive processes and generalized distributions. The reason is that in this case the operators mix under renormalization with total-derivative operators, implying that one now has to deal with an anomalous dimension matrix (ADM). Disentangling the elements of this matrix from the renormalization procedure is non-trivial. In this manuscript, we summarize two methods that can be used to reconstruct the anomalous dimensions.\newline

The article is organized as follows. In the next section, we briefly review some properties of the non-perturbative distributions and the operators in terms of which they are defined. Sec.~\ref{sec:ADM} then discusses two different methods to reconstruct the functional form of the operator anomalous dimensions. The first uses a consistency relation for the anomalous dimensions while the second will be based on conformal symmetry arguments. Besides employing different physical principles, these two methods also differ in the basis they use for the total-derivative operators. Hence the results of these methods can be compared by performing a basis transformation, which is briefly discussed at the end of Sec.~\ref{sec:ADM}.

\section{GPDs and composite operators}
\label{sec:operators}
Historically, GPDs were independently introduced in '94 by M\"uller \cite{Muller:1994ses} and '96-'97 by Radyushkin \cite{Radyushkin:1996nd} and Ji \cite{Ji:1996nm}. They can be thought of as generalizations of other types of non-perturbative QCD quantities like PDFs, form factors and distribution amplitudes. Physically they contain information about the transverse positions of partons inside hadrons. Furthermore, they give access to the contributions of the partonic orbital angular momentum to the total hadronic spin. As such, they are of great interest for the study of hadron structure, see e.g.~\cite{Pasquini:2008xm,Kaiser:2012hu,Bacchetta:2016ccz}. Their detailed study will be an important point of focus for several future colliders and experiments, including at a future electron-ion collider \cite{Boer:2011fh,AbdulKhalek:2021gbh,Anderle:2021wcy}, the LHeC \cite{LHeCStudyGroup:2012zhm} and JLab after the 22 GeV upgrade \cite{Accardi:2023chb}.\newline

\begin{figure}
           \centering
           \includegraphics[width=0.7\textwidth]{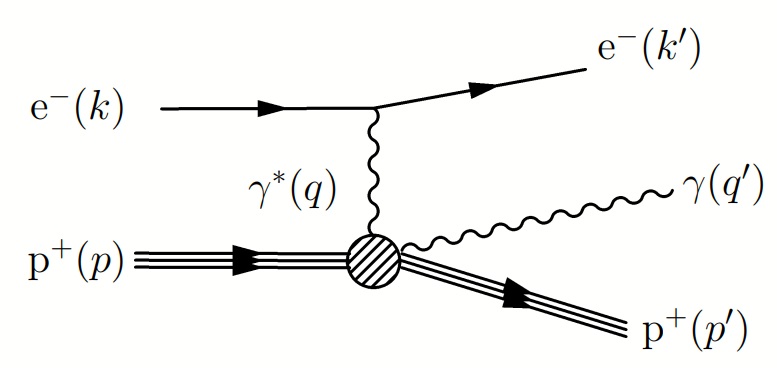}
           \caption{Sketch of the DVCS process.}
           \label{fig:DVCS}
\end{figure}

Experimentally, generalized distributions such as GPDs are accessible in hard exclusive processes, the cleanest example of which is deeply-virtual Compton scattering (DVCS), cf. Fig.~\ref{fig:DVCS}. 
Historically, the first measurements of this process were performed by HERMES \cite{HERMES:2001bob}, CLAS \cite{CLAS:2001wjj}, H1 \cite{H1:2001nez,H1:2005gdw} and ZEUS \cite{ZEUS:2003pwh} in the early 2000s. The process is characterized by four kinematic variables:
\begin{itemize}
    \item the virtuality $Q^2=-q^2$, which typically is taken to be large,
    \item Bjorken-x: $x_B=\frac{Q^2}{2p\cdot q}$
    \item the momentum transfer on the hadronic target: $t=(p-p')^2$ and
    \item the skewedness: $\xi=\frac{(p-p')^{+}}{(p+p')^{+}}$.
\end{itemize}
Here the skewedness is written in terms of lightcone coordinates
\begin{equation}
    p^{\pm}=\frac{1}{\sqrt{2}}(p^0\pm p^3).
\end{equation}
In the so-called (generalized) Bjorken limit, which amounts to taking $Q^2\rightarrow\infty$ while $x_B$ and $t$ are kept fixed, the DVCS amplitude factorizes into perturbative coefficient functions and non-perturbative GPDs. In particular, the coefficient functions correspond to partonic amplitudes while the GPDs correspond to hadronic matrix elements of composite QCD operators. The latter implies that the scale dependence of the distributions, which is vital information for phenomenological studies, is determined by the scale dependence of the operators. The distribution scale dependence is characterized by its evolution equation, which generically takes on the following form \cite{Muller:1994ses,Radyushkin:1996nd,Ji:1996nm}
\begin{equation}
\label{eq:GPVevol}
        \frac{\text{d}\mathcal{G}(x,\xi,t;\mu^2)}{\text{d}\ln\mu^2} = \int_{x}^{1}\frac{\text{d}y}{y}\mathcal{P}\Bigg(\frac{x}{y},\frac{\xi}{y}\Bigg)\mathcal{G}(y,\xi,t;\mu^2).
    \end{equation}
    Note that this is a generalization of the well-known DGLAP equation \cite{Gribov:1972ri,Altarelli:1977zs,Dokshitzer:1977sg}
\begin{equation}
    \frac{\text{d} f(x,\mu^2)}{\text{d} \ln{\mu^2}} = \int_x^1 \frac{\text{d}y}{y} P(y)f\Big(\frac{x}{y},\mu^2\Big).
\end{equation}
The evolution kernel $\mathcal{P}$ in Eq.(\ref{eq:GPVevol}) is directly related to the anomalous dimensions of the composite operators that define the distributions. Furthermore, contrary to the distributions, these anomalous dimensions can be computed perturbatively in QCD. In particular, they can be accessed through the renormalization of the \textit{partonic} operator matrix elements (OMEs). For the sake of explicitness we focus now on the leading-twist flavor non-singlet quark operators\footnote[2]{Note that higher-twist corrections are becoming increasingly important due to the high accuracy of experimental data and planned future experiments at lower energies. A recent overview of progress in higher-twist physics can be found in \cite{Braun:2022gzl}.}, which are defined as
\begin{equation}
        \mathcal{O} = \mathcal{S}\overline{\psi}\lambda^{\alpha}\Gamma D_{\mu_2}\dots D_{\mu_N}\psi.
\end{equation}
Here $D_{\mu} = \partial_{\mu}-i g_s A_{\mu}$ is the QCD covariant derivative and $\mathcal{S}$ denotes the fact that the operators are symmetrized in their Lorentz indices and traceless. This operation selects the leading-twist contributions of interest. Depending on the scattering process under consideration, different operators appear in the analysis. These are distinguished by their Dirac structure $\Gamma$. We fill focus our attention on the Wilson operators $\Gamma=\gamma_{\mu_1}$ and the transversity ones $\Gamma=\sigma_{\nu\mu_1}$. These operators are relevant for phenomenological studies of, for example, DVCS and transverse meson production. As we are interested in generalized distributions, for which exclusive processes need to be considered, the partonic OMEs need to be computed and renormalized in non-forward kinematics. This means that one needs to take into account mixing with total-derivative operators
{\begin{equation}
        \begin{pmatrix}
            \mathcal{O}_{N+1} \\ \partial \mathcal{O}_{N}\\ \vdots \\ \partial^N \mathcal{O}_{1}
    \end{pmatrix} 
    \,=\, 
    \begin{pmatrix}
            {Z_{N,N}} &  Z_{N,N-1} & ... &  Z_{N,0} \\
            0 & {Z_{N-1,N-1}} & ... &  Z_{N-1,0} \\
            \vdots & \vdots & ... & \vdots  \\
            0 & 0 & ...  &  {Z_{0,0}}
    \end{pmatrix} 
    \begin{pmatrix}
            [\mathcal{O}_{N+1}] \\ [\partial \mathcal{O}_{N}] \\ \vdots 
             \\ [\partial^N \mathcal{O}_{1}]
    \end{pmatrix}
    \end{equation}}
    which introduces a matrix of renormalization factors. As such, one also has an anomalous dimension matrix (ADM)
    {\begin{equation}
        \hat{\gamma} =-\frac{\text{d}\ln \hat{Z}}{\text{d}\ln\mu^2} = \begin{pmatrix}
            {\gamma_{N,N}} & \gamma_{N,N-1} & ... & \gamma_{N,0} \\
            0 & {\gamma_{N-1,N-1}} & ... & \gamma_{N-1,0} \\
            \vdots & \vdots & ... & \vdots  \\
            0 & 0 & ...  &  {\gamma_{0,0}}
    \end{pmatrix} .
    \end{equation}}
Note that the diagonal elements correspond to the forward anomalous dimensions, which determine the scale dependence of the standard PDFs. The reconstruction of the off-diagonal elements of this matrix from the renormalization procedure is non-trivial and will be described in the next section. In particular, we will review two independent methods based on (a) a consistency relation following from the renormalization structure of the operators and (b) conformal symmetry arguments.

\section{Reconstructing the anomalous dimensions}
\label{sec:ADM}
\subsection{Anomalous dimensions from a consistency relation}
By analyzing the renormalization structure of the operators and using relations between the total-derivative operators, one can derive the following consistency relation for the anomalous dimensions \cite{Moch:2021cdq}
\begin{equation}
\label{eq:relation}
    \gamma_{N,k}  = 
    \binom{N}{k}\sum_{j=0}^{N-k}(-1)^j \binom{N-k}{j}\gamma_{j+k,\:j+k}
    + \sum_{j=k}^N (-1)^k \binom{j}{k} \sum_{l=j+1}^N (-1)^l \binom{N}{l}  \gamma_{l,\:j}.
\end{equation}
This relation is valid to all orders in perturbation theory and can be used to reconstruct the functional form of the anomalous dimensions. In order to do so, one needs two ingredients. The first is the analytic structure of the forward anomalous dimensions, $\gamma_{N,N}$. These objects are known fully up to the three-loop level, with partial information being available up to five loops, see \cite{Falcioni:2022fdm,Gehrmann:2023ksf,Gross:1973ju,Gross:1974cs,Floratos:1977au,Gonzalez-Arroyo:1979guc,Floratos:1978ny,Gonzalez-Arroyo:1979qht,Gonzalez-Arroyo:1979kjx,Curci:1980uw,Furmanski:1980cm,Shifman:1980dk,Baldracchini:1981,Artru:1989zv,Gracey:1994nn,Gracey:1996ad,Hayashigaki:1997dn,Kumano:1997qp,Blumlein:1997bz,Larin:1996wd,Vogelsang:1997ak,Bennett:1997ch,Blumlein:1997bs,Blumlein:1998hz,Blumlein:1998pp,vanNeerven:2000wp,Blumlein:2001ca,Gracey:2003yr,Gracey:2003mr,Vogt:2004mw,Moch:2004pa,Blumlein:2004bb,Gracey:2006zr,Gracey:2006ah,Blumlein:2009tj,Bierenbaum:2009mv,Vogt:2010ik,Soar:2009yh,Vogt:2010pe,Ablinger:2010ty,Velizhanin:2011es,Velizhanin:2012nm,Ablinger:2014vwa,Ablinger:2014nga,Moch:2014sna,Ruijl:2016pkm,Davies:2016jie,Moch:2017uml,Ablinger:2017tan,Vogt:2018miu,Moch:2018wjh,Behring:2019tus,Herzog:2018kwj,Velizhanin:2014fua,Blumlein:2021enk,Blumlein:2021ryt,Moch:2021qrk,Blumlein:2022kqz,Blumlein:2023aso,Gehrmann:2023cqm,Gehrmann:2023iah,Falcioni:2023luc,Ji:2023eni,Falcioni:2023vqq,Falcioni:2023tzp,Moch:2023tdj}\footnote[3]{In the large-$n_f$ limit the all-order expressions are known, see \cite{Gracey:1994nn,Gracey:2003mr}. The generalization of these results to non-forward kinematics can be found in \cite{VanThurenhout:2022hgd}.}. The second ingredient consists of a boundary condition, which needs to be supplied in order to ensure the uniqueness of the result. The r\^ole of this boundary condition is played by the last column of the ADM, $\gamma_{N,0}$, which can be directly related to the computation of the appropriate OMEs. We refer the reader to \cite{Moch:2021cdq} for more details. The Feynman rules for the quark operator vertices, which are of course necessary for the computation of the OMEs, were derived in \cite{Mikhailov:2020tta} in x-space and independently in \cite{Somogyi:2024njx} in Mellin space. The rules presented in the latter are also implemented in {\sc Form} and {\sc Mathematica}, available at \url{https://github.com/vtsam/NKLO}. They are valid for quark operators with an arbitrary number of total derivatives and applicable to any order in perturbation theory. Applications of the consistency relation in the large-$n_f$ and the planar limits can be found in \cite{Moch:2021cdq,VanThurenhout:2022nmx}. The main advantage of the method is that the full procedure can be easily automated using standard computer algebra methods, meaning that it is, at least in principle, straightforward to apply at higher orders in perturbation theory.

\subsection{Anomalous dimensions from conformal symmetry}
Another approach to reconstruct the anomalous dimensions is based on conformal symmetry arguments \cite{Ji:2023eni,Efremov:1978rn,Balitsky:1987bk,Mueller:1991gd,Belitsky:1998gc,Braun:2016qlg,Braun:2017cih,Braun:2021tzi,Braun:2022byg}. Instead of working with physical QCD in four dimensions, one considers QCD in $D=4-2\eps$ dimensions at the critical point. Here the QCD beta-function vanishes, and hence QCD becomes inherently conformal. The operator anomalous dimensions $\gamma^{\mathcal{C}}$ can then be reconstructed using consistency relations coming from the conformal algebra. The physical evolution kernels have the same functional form as the critical ones, up to terms associated to the explicit breaking of conformal symmetry. This is associated to the QCD beta-function and the so-called conformal anomaly. The latter can be computed perturbatively and is currently known to two-loop accuracy \cite{Mueller:1991gd,Braun:2016qlg,Braun:2017cih}. As generically the anomalous dimensions depend only on the
conformal anomaly at one order in perturbation theory lower \cite{Mueller:1991gd}, this method has been successfully applied to compute them up to three loops \cite{Braun:2017cih}. A nice advantage of this method is that, because of exact conformal symmetry at leading order, the one-loop ADM is actually diagonal in this set-up, while the main bottleneck is the (automated) computation of the anomaly.

\subsection{Connecting results}
Note that the two approaches discussed above to reconstruct the anomalous dimensions are quite different, being based on completely different physical ideas. Furthermore, in practice, they use different bases for the total-derivative operators. As such, one can compare the results of both methods by setting up the appropriate basis transformation. The derivation of this transformation formula was presented in \cite{VanThurenhout:2023gmo}, and we just cite the result here
     \begin{align}
     \label{eq:basisTrans}
     \begin{split}
       \gamma_{N,k}^{\D} &= 
    \frac{(-1)^k(N+1)!}{(k+1)!}\sum_{l=k}^{N}(-1)^l\binom{N}{l}\frac{l!(3+2l)}{(N+l+3)!}\sum_{j=k}^{l}\binom{j}{k}\frac{(j+k+2)!}{j!}\gamma_{l,j}^{\mathcal{C}}, \\
    \gamma_{N,k}^{\mathcal{C}} &= 
    (-1)^k\frac{k!}{N!}(3+2k)\sum_{l=k}^{N}(-1)^{l}\binom{N}{l}\frac{(N+l+2)!}{(l+1)!}\sum_{j=k}^{l}\binom{j}{k}\frac{(j+1)!}{(j+k+3)!}\gamma_{l,j}^{\D}.
    \end{split}
    \end{align}
Here $\gamma^{\mathcal{D}}$ represents the anomalous dimensions in the operator basis used to derive the consistency relation in Eq.~(\ref{eq:relation}), while $\gamma^{\mathcal{C}}$ are written in the conformal basis. With these explicit expressions, one can cross-check a priori independent computations. Furthermore, they can be used to learn about the functional form of the ADM in one basis if it is already known in the other one.\newline

Both methods come with their own advantages and can be used in a complementary fashion. For example, at the two-loop level, the leading-color part of the ADM is straightforwardly computed using Eq.~(\ref{eq:relation}). The subleading-color terms however are non-trivial because of the non-trivial structure of the corresponding forward anomalous dimensions. Interestingly, the situation is reversed in the conformal approach, as here the subleading-color part is straightforward since it actually does not depend on the conformal anomaly. The leading-color part however does get contributions from the anomaly. Similar comments hold also at higher orders in perturbation theory. As such, one way to proceed is to compute different contributions to the anomalous dimensions in different bases and then use Eq.~(\ref{eq:basisTrans}) to write the results in the same operator basis.

\section{Summary}
\label{sec:conclusion}
We have discussed two methods to reconstruct the anomalous dimensions of composite operators in non-forward kinematics using (a) a consistency relation and (b) conformal symmetry arguments. These anomalous dimensions determine the scale dependence of non-forward parton distributions, and hence are of importance for phenomenological studies of hard hadronic scattering processes. Both methods come with their own advantages, and can be used in a complementary fashion in the determination of the anomalous dimensions.

\subsection*{Acknowledgements}
Part of this work has been supported by grant K143451 of the National Research, Development and Innovation Fund in Hungary.

\bibliographystyle{JHEP}
\bibliography{omebib}

\end{document}